\documentclass[12pt,preprint]{aastex}
\usepackage{epsf}

\shorttitle{Distribution Function of Phase Sum}
\shortauthors{Hikage, Matsubara, \& Suto}
\newcommand{\mb}[1]{\mbox{\boldmath $#1$}}
\newcommand{\smb}[1]{\mbox{\scriptsize\boldmath $#1$}}
\begin{document}
\title{The Distribution Function of the Phase Sum as a Signature of
Phase Correlations Induced by Nonlinear 
Gravitational Clustering}
\author{Chiaki Hikage\altaffilmark{1}, Takahiko Matsubara\altaffilmark{2} 
and Yasushi Suto\altaffilmark{1,3}}
\altaffiltext{1}{Department of Physics, School of Science, University of
    Tokyo, Tokyo 113-0033, Japan.}
\altaffiltext{2}{Department of Physics and Astrophysics, 
Nagoya University, Chikusa, Nagoya 464-8602}
\altaffiltext{3}{Research Center for
    the Early Universe (RESCEU), School of Science, University of
    Tokyo, Tokyo 113-0033, Japan.}
\email{hikage@utap.phys.s.u-tokyo.ac.jp, taka@a.phys.nagoya-u.ac.jp, 
suto@phys.s.u-tokyo.ac.jp}
\received{2003 August 11}
\accepted{2003 ???}
\begin{abstract}
We explore a signature of phase correlations in Fourier modes of dark
matter density fields induced by nonlinear gravitational
clustering. We compute the distribution function of the phase sum of
the Fourier modes, $\theta_{{\smb k}_1}+\theta_{{\smb
k}_2}+\theta_{{\smb k}_3}$, for triangle wavevectors satisfying ${\mb
k}_1+{\mb k}_2+{\mb k}_3={\mb 0}$, and compare with the analytic
prediction in perturbation theory recently derived by one of us. Using
a series of cosmological $N$-body simulations, we extensively examine
the time evolution and the dependence on the configuration of
triangles and the sampling volume. Overall we find that the numerical
results are remarkably consistent with the analytic formula from the
perturbation theory. Interestingly the validity of the perturbation
theory at a scale $k$ corresponding to the wavevector ${\mb k}$ is
determined by $P(k)/V_{\rm samp}$, the ratio of the power spectrum
$P(k)$ and the sampling volume $V_{\rm samp}$, not by $k^3P(k)$ as in
the case of the conventional cosmological perturbation theory.
Consequently this statistics of phase correlations is sensitive to the
size of the sampling volume itself. This feature does not show up in
more conventional cosmological statistics including the one-point
density distribution function and the two-point correlation functions
except as a sample-to-sample variation.  Similarly if the sampling
volume size $V_{\rm samp}$ is fixed, the stronger phase correlation
emerges first at the wavevector where $P(k)$ becomes largest, i.e.,
{\it in linear regimes according to the standard cosmological
perturbation theory}, while the distribution of the phase sum stays
fairly uniform in nonlinear regimes.  The above feature can be
naturally understood from the corresponding density structures in real
space as we discuss in detail.
\end{abstract}

\section{Introduction \label{sec:intro}}

The most conventional statistics in cosmology include the two-point
correlation function in real space and the power spectrum in Fourier
space. They have been investigated both by numerous analytical and
numerical methods and provided useful insights into the cosmological
parameters and properties of the galaxy biasing among others mainly
from the large-scale structure of the universe.  The present universe,
however, has much more complex structures, such as filaments, voids,
and superclusters, which cannot be fully described by the two-point
statistics.  Therefore other statistics beyond the two-point
statistics are desperately required for better understanding of the
universe.

The Fourier transform $f_{\smb k}$ of the density fluctuations,
$\delta({\mb x})=(\rho({\mb x})-\bar{\rho})/\bar{\rho}$, where
$\rho({\mb x})$ is the density field and $\bar{\rho}$ is the spatial
mean of the density field, is expressed in terms of the modulus
$|f_{\smb k}|$ and the phase $\theta_{\smb k}$ as follows:
\begin{equation}
f_{\smb k}=|f_{\smb k}|\exp(i\theta_{\smb k}).
\end{equation}
The two-point statistics is defined in terms of the modulus alone, and
thus the statistics of phases play complementary roles in
characterizing the nature of the density field. Especially, the
correlation of phases among different Fourier modes is a key
ingredient in understanding the emergence of the non-Gaussian
signature from the primordial Gaussian density field.

In fact several statistics which carry the phase information have been
already proposed in cosmology, including the void probability function
\citep{White1979}, the genus statistics \citep{GMD1986}, the Minkowski
functionals \citep{MBW1994,SB1997}.  However, finding useful
statistics of the Fourier phase itself is difficult mainly due to the
cyclic property of the phase.  For example, the one-point phase
distribution turns out to be essentially uniform even in a strongly
non-Gaussian field \citep{Suginohara1991} and then one cannot extract
any useful information out of it. For this reason, previous studies of
the Fourier phase have been mainly devoted to the evolution of phase
shifts in individual modes \citep{Ryden1991, Soda1992, Jain1998}), and
the phase differences between the Fourier modes \citep{SMS1991,
CC2000, C2001, CCN2002, WCM2003}.  There is still, however, a very
incomplete understanding of how phase correlations among different
modes start to show up, or for the corresponding structure in real
space that the strong phase correlation indicates.

The connection between the higher-order statistics and the phase
correlations has been also suggested \citep{Bertschinger1992, WC2003}.
Recently, one of the present authors obtained an analytic expression
for the distribution function of the ``phase sum'' $\theta_{{\smb
k}_1}+\theta_{{\smb k}_2}+\cdots + \theta_{{\smb k}_N}$ where the
corresponding wavevectors satisfy ${\mb k}_1+{\mb k}_2+ \cdots + {\mb
k}_N={\mb 0}$ (Matsubara 2003b).  He discovered the general relation
between the distribution of the phase sum and the hierarchy of
polyspectra in the perturbation theory. Following his analytic
results, we extensively study the behavior of the distribution of the
phase sum of triangle wavevectors using a series of cosmological
$N$-body simulation.

The plan of this paper is as follows. In \S 2, we briefly review the
analytic expansion formula of the phase correlations with particular
emphasis to the sampling volume dependence on the distribution
function of the phase sum. Section 3 summarizes the simulation data
and the computation method, and the results are shown in \S 4.
Finally \S 5 is devoted to conclusions and discussion.

\section{Series expansion of the Distribution of the Phase Sum}

Starting from the Edgeworth-like expansion of the joint probability
distribution function (PDF) \citep{Matsubara1995,Matsubara2003a} of
arbitrary sets of the modulus factor $|f_{\smb k}|$ and the phase
factor $\theta_{\smb k}$ in the Fourier mode ${\mb k}$, Matsubara
(2003b) derived the joint PDF of phases among different Fourier modes
of arbitrary closed wavevectors after integrating over the other
variables. The condition of the closed wavevectors ensures the
translational invariance of the statistics. In the lowest order
approximation, the joint PDF of the three phases in triangle
wavevectors $\theta_{{\smb k}_1}, \theta_{{\smb k}_2}$ and
$\theta_{{\smb k}_3}=-\theta_{{\smb k}_1+{\smb k}_2}$ (${\mb
k}_3=-{\mb k}_1-{\mb k}_2$, ${\mb k}_1 \neq {\mb k}_2$) in a sampling
volume $V_{\rm samp}$ is written \citep{Matsubara2003b} as
\begin{equation}
\label{eq:jointPDF}
{\cal P}(\theta_{{\smb k}_1},\theta_{{\smb k}_2},\theta_{{{\smb k}_1}
+{{\smb k}_2}}| V_{\rm samp})\propto
1+\frac{\pi^{3/2}}{4}p^{(3)}({{\mb k}_1},{{\mb k}_2}|V_{\rm samp})
\cos(\theta_{{\smb k}_1}+ \theta_{{\smb k}_2}-\theta_{{{\smb
k}_1}+{{\smb k}_2}}),
\end{equation}
\begin{equation}
\label{eq:p3}
p^{(3)}({\mb k}_1, {\mb k}_2|V_{\rm samp}) = \frac{B({\mb k}_1, {\mb
k}_2)}{\sqrt{V_{\rm samp} P(|{\mb k}_1|)P(|{\mb k}_2|)P(|{\mb
k}_1+{\mb k}_2|)}}.
\end{equation}
For any periodic distributions on $[0,2\pi]$, it is natural to expect
that its lowest-order expansion gives a constant plus the cosine term.
Therefore the most important point in the above formula lies in the
fact that the proportional factor $p^{(3)}$ is given by the
cumulant defined in terms of the bispectrum $B({\mb k}_1,{\mb k}_2)$
and the power spectrum $P(k)$ as follows:
\begin{eqnarray}
\langle f_{{\smb k}}f_{{\smb k}^\prime} \rangle & = & \delta^K_{{\smb
k}+{\smb k}^\prime} P(|{\mb k}|), \\ \langle f_{{\smb k}_1}f_{{\smb
k}_2}f_{{\smb k}_3} \rangle & = & V_{\rm samp}^{-1/2} \delta^K_{{\smb
k}_1+{\smb k}_2+{\smb k}_3} B({\mb k}_1,{\mb k}_2),
\end{eqnarray}
where $\langle\cdot\cdot\cdot\rangle$ is the ensemble average and
$\delta^K_{\smb k}$ denotes the Kronecker delta which is $1$ only when
${\mb k} = {\mb 0}$. We use the convention of the Fourier transform
$f_{\smb k} = V^{-1/2} \int_V d^3x e^{-i{\smb k}\cdot{\smb x}} f({\mb
x})$ in a finite box-size $V$.  We explicitly write the volume
dependence of $p^{(3)}$, which is derived in Appendix
\ref{app:extensivity}.  Equation (\ref{eq:jointPDF}) implies that the
PDF of the phase sum, $\theta_{{\smb k}_1}+\theta_{{\smb k}_2} -
\theta_{{{\smb k}_1}+{{\smb k}_2}}$, is written as
\begin{equation}
\label{eq:matsubara}
{\cal P}(\theta_{{\smb k}_1}+\theta_{{\smb k}_2}
-\theta_{{{\smb k}_1}+{{\smb k}_2}}|V_{\rm samp}) \propto
1+\frac{\pi^{3/2}}{4}p^{(3)}({\mb k}_1,{\mb k}_2|V_{\rm samp})
\cos (\theta_{{\smb k}_1}+\theta_{{\smb k}_2}
-\theta_{{{\smb k}_1}+{{\smb k}_2}}).
\end{equation}
We assume the statistical isotropy in the field distribution in the
following analysis and thereby the distribution of the phase sum
depends on $|{\mb k}_1|$, $|{\mb k}_2|$ and $\theta_{12}$ in a fixed
box-size $V_{\rm samp}$.

If the hierarchical clustering ansatz is valid, $B({\mb k}_1,{\mb
k}_2) \sim P({\mb k}_1)P({\mb k}_2)$ and thereby $p^{(3)}$ is
approximately given by $\sqrt{P(k)/V_{\rm samp}}$. Therefore, the
lowest-order approximation (eq.[\ref{eq:matsubara}]) breaks down at a
scale $k$ satisfying the condition $P(k)/V_{\rm samp} > 1$, which is
different from the condition of the nonlinear clustering, $k^3 P(k) >
1$. When $V_{\rm samp}$ is sufficiently large, the higher-order terms
$p^{(N)}$ become negligible and equation (\ref{eq:matsubara}) is
applicable even in nonlinear regimes in the sense of the conventional
cosmological perturbation theory.

The unusual feature of the statistics of the phase sum is that the
degree of the non-uniformity explicitly depends on the size of the
sampling volume $V_{\rm samp}$. This comes from the fact that the
$N$th-order terms $p^{(N)}$ have different volume dependence, $
\propto V_{\rm samp}^{1-N/2}$ (see Appendix \ref{app:extensivity}).
Therefore, the distribution of phases in an infinite volume is always
random in a statistical sense. The concept of the phase distribution
function is meaningful only when averaged over many subsamples of a
fixed $V_{\rm samp}$ within the whole sample.

\section{$N$-body Simulation Data and Computational Method}

\begin{deluxetable}{ccccc}
\footnotesize
\tablecaption{Simulation model parameters: the dimensionless
matter-density parameter $\Omega_0$; the dimensionless cosmological 
constant $\lambda_0$; the shape parameter $\Gamma$ of the CDM transfer
function (Bardeen et al. 1986); the r.m.s. density fluctuation amplitude 
smoothed by top-hat filter with the scale of $8h^{-1}$Mpc, $\sigma_8$.}
\tablewidth{0pt}
\tablehead{
  \colhead{Model} & \colhead{$\Omega_0$} & \colhead{$\lambda_0$} 
  & \colhead{$\Gamma$} & \colhead{$\sigma_8$}
}
\startdata
  LCDM & $0.3$ & $0.7$ & $0.21$ & $1$  \\ 
  SCDM & $1$   & $0$   & $0.5$  & $0.6$ \\ 
  OCDM & $0.3$ & $0$   & $0.25$ & $1$  \\
  scale-free ($n=-2,-1,0,+1$) & $1$ & $0$ & -- & -- \\ 
\enddata
\label{tab:modelpara}
\end{deluxetable}

\begin{deluxetable}{ccc}
\footnotesize
\tablecaption{The original simulation box-size $V_{\rm box}$ and
the ratio of sampling volume to original box-size, $V_{\rm samp}/V_{\rm box}$.}
\tablewidth{0pt}
\tablehead{
   \colhead{Model} & \colhead{$V_{\rm box}$} 
  & \colhead{$V_{\rm samp}$ / $V_{\rm box}$}
}
\startdata
  CDM (LCDM,SCDM,OCDM) & ($300h^{-1}$Mpc)$^3$ & 1 \\
  CDM (LCDM,SCDM,OCDM) & ($100h^{-1}$Mpc)$^3$ & 1, 1/8, 1/64 \\
  scale-free ($n=-2,-1,0,+1$) & -- & 1, 1/8, 1/64 \\
\enddata
\label{tab:boxsize}
\end{deluxetable}

In order to study the behavior of the distribution of the phase sum, we
apply the analytic formula (eq.[\ref{eq:matsubara}]) on dark matter
distribution using a series of P${}^3$M $N$-body simulations 
(Jing 1998; Jing \& Suto 1998).
These simulations employ $256^3$ particles in a cubic box
with $256^3$ meshes and start from Gaussian-random distribution with
periodic boundary conditions. We adopt a variety of the initial power
spectra including four scale-free models with power-law indices $n=-2,
-1, 0$, and $+1$. The simulations were terminated at the scale factor
$a=1.0$ when the r.m.s. mass fluctuation smoothed over the top-hat
filter with the scale of one-tenth of a box-length reached unity.  We
use the scale-free models at different scale factors 
so as to study the evolution of the distribution of
the phase sum.  We also analyze three cold dark matter (CDM) models,
Lambda CDM (LCDM), Standard CDM (SCDM), and Open CDM(OCDM) at $z=0$.
The model parameters of all above simulations are summarized in 
Table \ref{tab:modelpara}.

We analyze three different realizations for almost all of scale-free
models and CDM models to estimate the sample-to-sample variance. In
order to study the box-size dependence of phase correlations, we
consider the CDM data with different box-size $V_{\rm box}$ of
$(300h^{-1}{\rm Mpc})^3$ and $(100h^{-1}{\rm Mpc})^3$. Furthermore, we
divide the original box data based on all of scale-free models and CDM
models with volume of $(100h^{-1}{\rm Mpc})^3$ into several sub-boxes
with an equal volume $V_{\rm samp}$: $8$ cubic boxes with a half length, 
and $64$ cubic boxes with a quarter length of the original box, 
as shown in Table \ref{tab:boxsize}.

The computation of the PDF of the phase sum proceeds as follows: first
we construct the density fields defined on grid points using the
cloud-in-cell interpolation from particle distribution, and then
Fourier-transform them into $k$-space. Next we choose two wavevectors
${\mb k}_1$ and ${\mb k}_2$ whose absolute values and open angle are
within a certain range. In what follows, we consider, for simplicity and
definiteness, only the configuration of $|{\mb k}_1|$ and $|{\mb k}_2|$
whose absolute values are within the same range. Finally we obtain three
phases of the Fourier modes for all sets of the chosen wavevectors,
${\mb k}_1$, ${\mb k}_2$ and ${\mb k}_1+{\mb k}_2$ and then compute the
distribution of the phase sum $\theta_{{\smb k}_1}+\theta_{{\smb
k}_2}-\theta_{{\smb k}_1+ {\smb k}_2}$ divided by the mean number per
binned phase sum. We set a bin width of $0.1\pi$ in $2\pi$ range.  We
estimate $p^{(3)}$ in two ways; one is based on a fit of the PDF of the
phase sum to the analytic formula of equation (\ref{eq:matsubara}), and
the other uses the direct definition (eq.[\ref{eq:p3}]) in terms of the
power spectrum and the bispectrum averaged over the same range of $|{\mb
k}_1|$, $|{\mb k}_2|$ and $\theta_{12}$.

Throughout the paper the modulus of each wavevector is expressed in
units of the Nyquist wavenumber of the grid in each field unless the
units of the scale are explicitly mentioned. For instance, $k=[a,b]$
implies that the two wavevectors satisfy $a< |\mb k_1|/k_{\rm Nyquist}
<b$, and $a< |\mb k_2|/k_{\rm Nyquist} <b$.  The Nyquist wavenumber is
defined by
\begin{equation}
k_{\rm Nyquist}=\frac{2\pi}{V_{\rm samp}^{1/3}}\frac{n_{\rm grid}}{2},
\end{equation}
where $n_{\rm grid}$ is the number of grids in a length of each box and
$V_{\rm samp}^{1/3}$ is the length of the sampling box.  In scale-free
models, we choose $n_{\rm grid}=256\times(V_{\rm samp}/V_{\rm box})^{1/3}$ 
so as to fix $k_{\rm Nyquist}$ as
$128 \times 2\pi/V_{\rm box}^{1/3}$, where $V_{\rm box}$ is the original
simulation volume.  We also confirm that the choice of $n_{\rm grid}$
does not change the results which we present in the next section.

\section{Results}

\subsection{Dependence of the phase sum PDF on the nonlinear
  gravitational evolution, scale and angle of the wavevectors, and the
sampling volume size}

\begin{figure*}[tph]
\begin{center}
\epsfxsize=16cm
\plotone{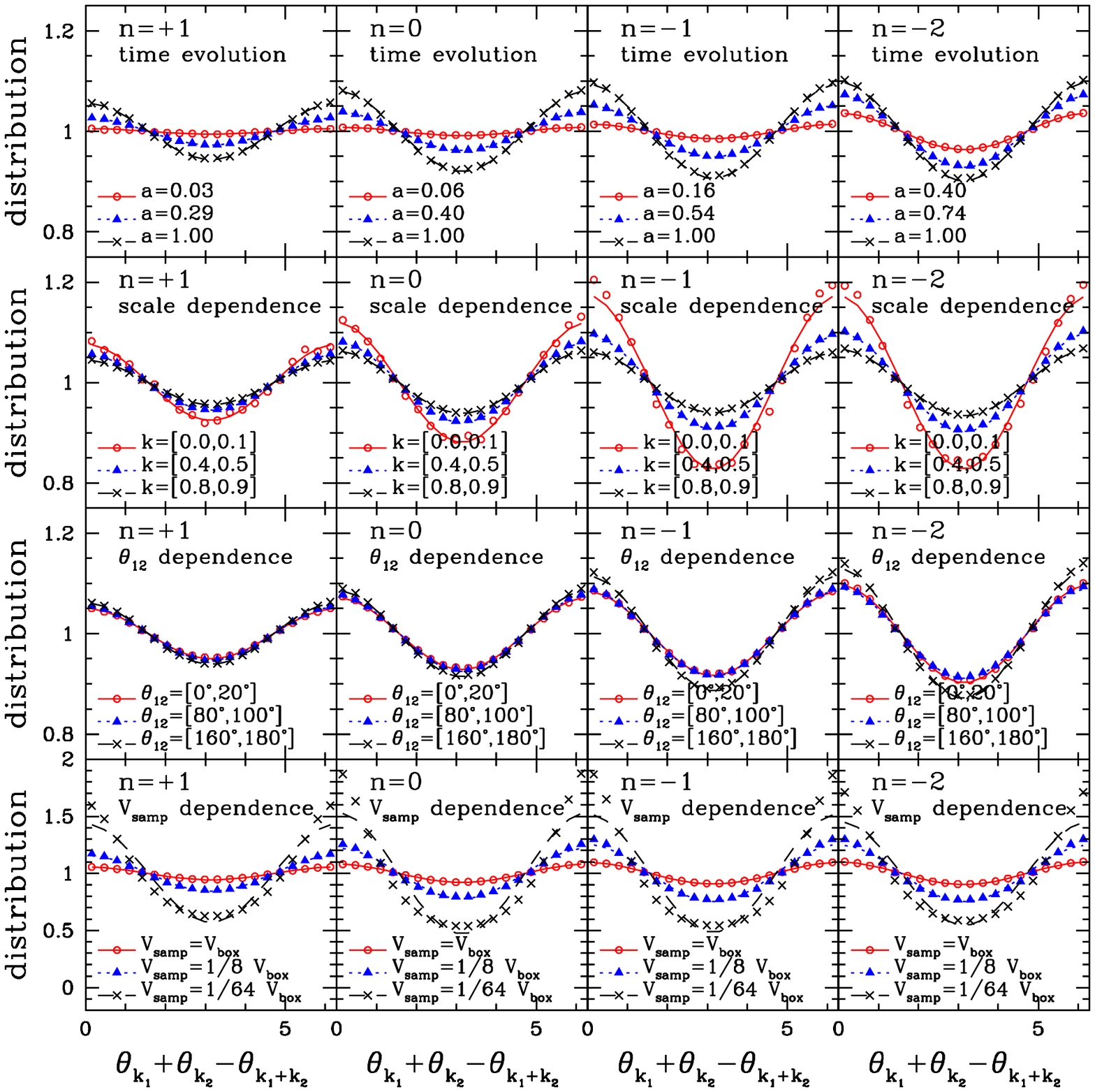}
\end{center}
\caption{Various parameter dependence of the distribution of 
the phase sum $\theta_{{\smb k}_1}+\theta_{{\smb k}_2}-\theta_{{\smb k}_1
+{\smb k}_2}$ plotted by symbols in comparison with the perturbative formula 
(eq.[\ref{eq:matsubara}]) plotted by lines for scale-free models. 
The power-law index $n$ decreases from left to right.
Each panel shows the time evolution, the scale dependence, 
the $\theta_{12}$ dependence, and the sampling volume $V_{\rm samp}$ 
dependence from top to bottom.
\label{fig:phasesum_scalefree}}
\end{figure*}

Figure \ref{fig:phasesum_scalefree} shows the time evolution, and the
dependence on $k$, $\theta_{12}$ and $V_{\rm samp}$, of the
distribution function of the phase sum (symbols) for scale-free models
with power-law indices $n=-2$, $-1$, $0$, and $+1$. For comparison, we
plot with lines the lowest-order analytic formula
(eq.[\ref{eq:matsubara}]) with $p^{(3)}$ directly evaluated from power
spectra and bi-spectra of simulations. We note that the Poisson error
of the distribution of the number of binned phase sum is very small.

\begin{figure*}[tph]
\begin{center}
\epsfxsize=16cm 
\plotone{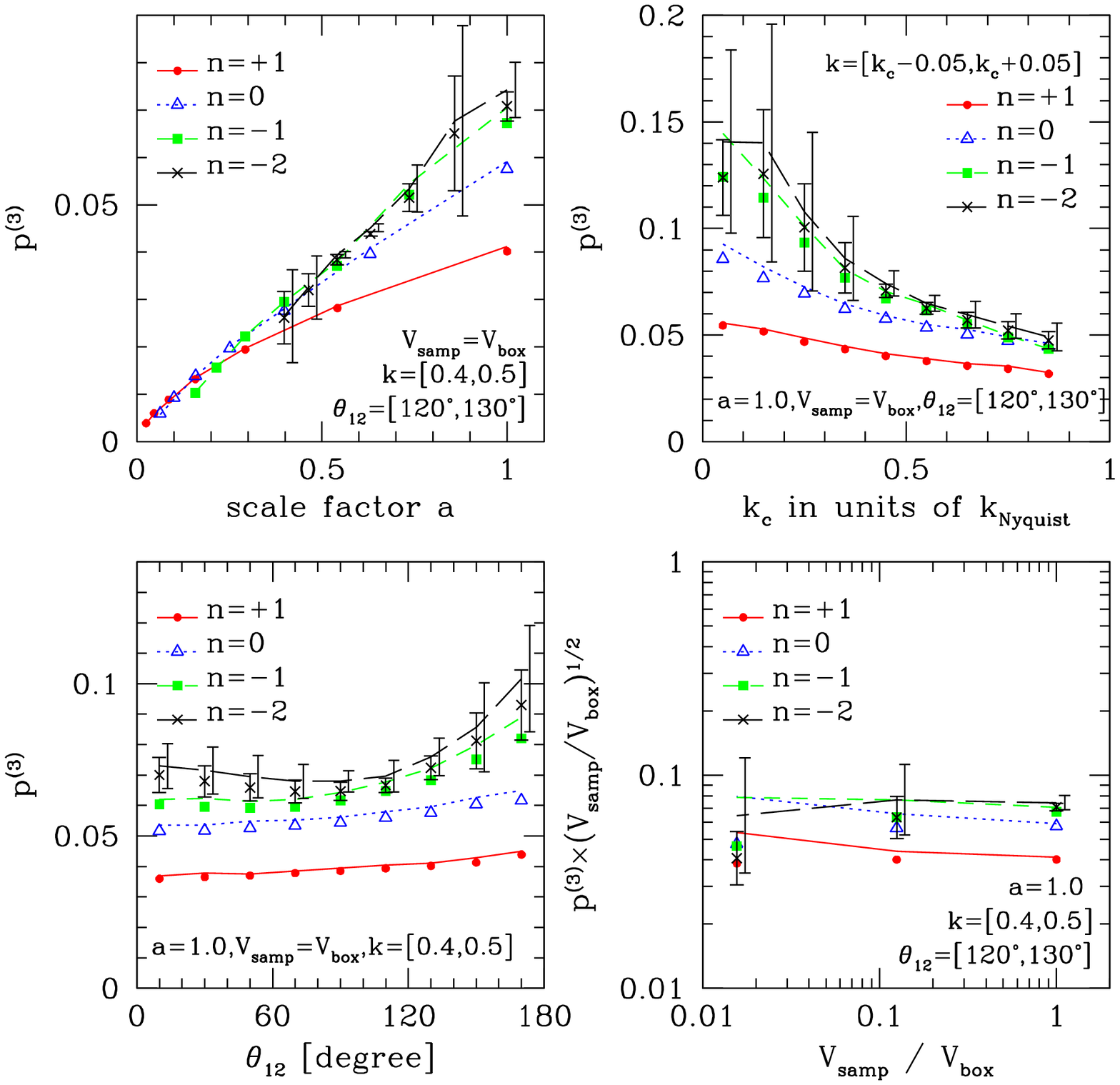}
\end{center}
\caption{Comparison of $p^{(3)}({\mb k}_1,{\mb k}_2)$ plotted by symbols with
the amplitude of the distribution of phase sum 
at $\theta_{{\smb k}_1}+\theta_{{\smb k}_2}-\theta_{{\smb k}_1+{\smb k}_2}=0$
divided by the constant factor $\pi^{3/2}/4$ (see eq.[\ref{eq:matsubara}])
plotted by lines for scale free models. 
The error-bars indicate the sample variances of $p^{(3)}$ and
the amplitude of the distribution of phase sum for $n=-2$ model.
Each panel shows the dependence of scale factor $a$ ({\it Upper-Left}),
scale({\it Upper-Right}),  $\theta_{12}$ ({\it Lower-Left}) 
and sampling volume $V_{\rm samp}$ ({\it Lower-Right}).
\label{fig:p3_scalefree}}
\end{figure*}

\begin{figure*}
\begin{center}
\epsfxsize=16cm 
\plotone{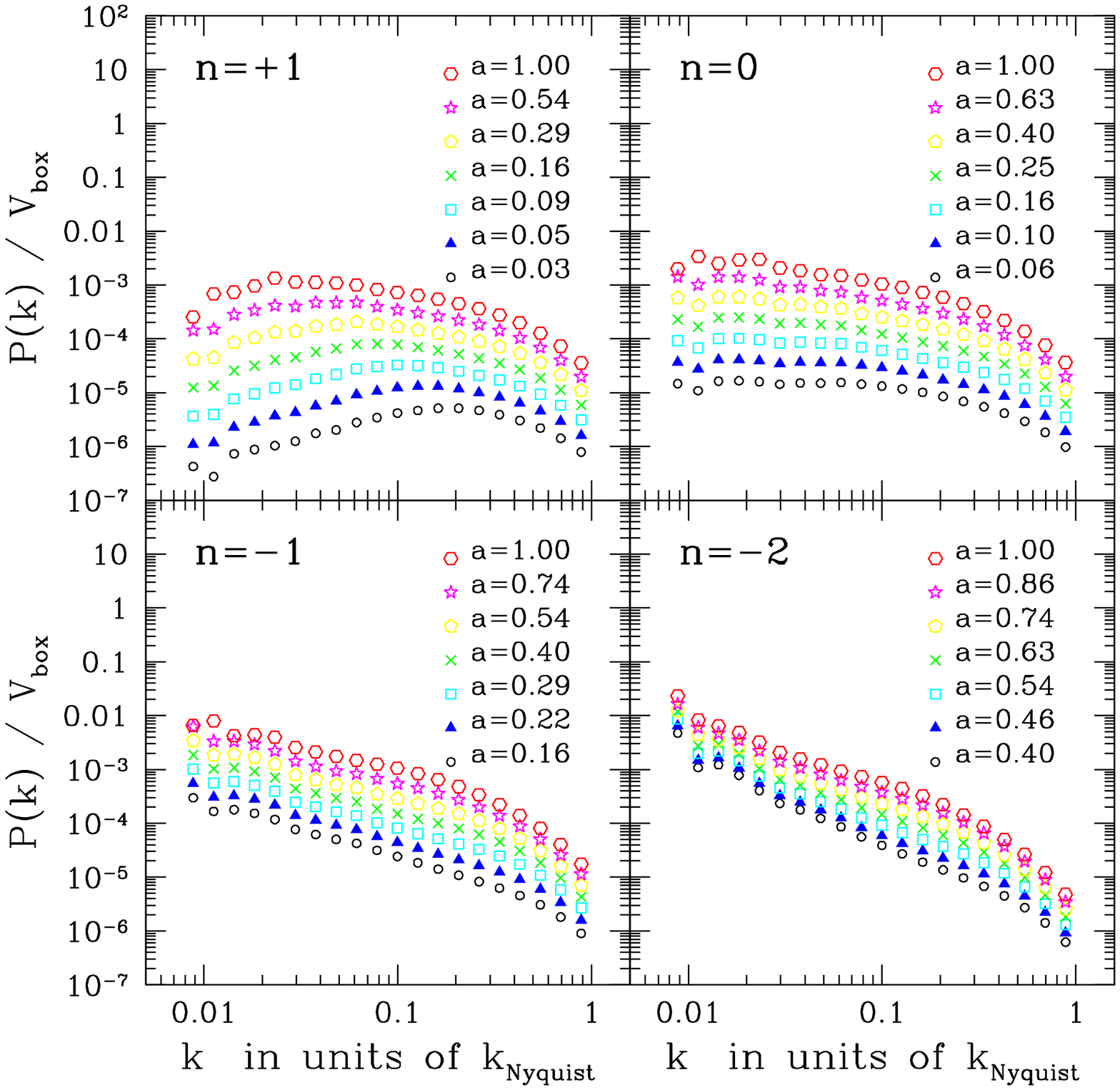}
\end{center}
\caption{Power spectrum $P(k)$ in units of $V_{\rm box}$ 
at various epochs in scale free models
with $n=1$({\it Upper-left}), $n=0$({\it Upper-right}), 
$n=-1$({\it Lower-left}) and $n=-2$({\it Lower-right}).
\label{fig:power_scalefree}}
\end{figure*}

 The symbols in Figure \ref{fig:p3_scalefree} plot the cumulant
$p^{(3)}$ as a function of the scale factor $a$, the central value of
the wavevector $k_{\rm c}$ ($k=[k_{\rm c}-0.05, k_{\rm c}+0.05]$),
$\theta_{12}$, and $V_{\rm samp}$. The corresponding lines indicate
$p^{(3)}$ evaluated from a fit of the phase-sum distribution to the 
analytic formula of equation (\ref{eq:matsubara}). 
We choose the fitting range as $\theta_{{\smb k}_1}+ 
\theta_{{\smb k}_2}-\theta_{{\smb k}_1+{\smb k}_2} =[1.9\pi,2\pi]
$ and $[0\pi,0.1\pi]$ where the deviation from the lowest-order 
approximation is clearly seen.
The quoted error-bars indicate the sample variance for $n=-2$ model.

Figures \ref{fig:phasesum_scalefree} and \ref{fig:p3_scalefree} employ a
fiducial set of parameters,  $a=1$, $k=[0.4,0.5]$,
$\theta_{12}=[120^\circ, 130^\circ]$, and $V_{\rm samp}=V_{\rm box}$,
except for one parameter that is examined in each panel.

The top panels in Figure \ref{fig:phasesum_scalefree} show the
evolution of the PDF of the phase sum. In the earliest epochs, the
density fields are still close to Gaussian and the PDF of the phase
sum remains almost uniform.  While the non-uniformity of the PDF
develops at subsequent epochs, the upper-left panel in Figure
\ref{fig:p3_scalefree} indicates that the lowest-order perturbation
result well agrees with the simulation data even up to $a=1$.

The second-upper panels in Figure \ref{fig:phasesum_scalefree} show
the scale dependence of the phase-sum PDF. Again the overall agreement
between the lowest-order perturbation theory and simulation data is
very good.  Nevertheless only on large scales ($k=[0.0, 0.1]$), the
lowest-order approximation starts to become invalid (see also the
upper-right panel of Fig.\ref{fig:p3_scalefree}).  The fluctuations on
large scales should be closer to the linear stage, and thus this
behavior seems to be inconsistent with a naive expectation that
non-Gaussianity should show up first from small scales. This apparent
discrepancy can be reconciled by the fact that the cumulant $p^{(3)}$
is approximately proportional to $\sqrt{P(k)/V_{\rm samp}}$ as we
emphasized in \S 2.  In all the current data of the scale-free models,
the nonlinearly evolved power spectra $P(k)$ decrease with $k$
(Fig. \ref{fig:power_scalefree}). Therefore, as long as $V_{\rm samp}$
is fixed, $p^{(3)}$ is {\it smaller for larger $k$}, and thus the
lowest-order perturbation provides a better approximation. This is an
interesting and unique feature of the phase sum statistics that we
examine here.

The second-lower panels in Figure \ref{fig:phasesum_scalefree} show the
$\theta_{12}$ dependence of the phase-sum distribution.  As indicated
more clearly in the lower-left panel of Figure \ref{fig:p3_scalefree},
the value of $p^{(3)}$, and thus the non-uniformity of the phase-sum
PDF, increases slightly at $\theta_{12}\ge 120^\circ$ for $n=-1$ and
$-2$ models. We suspect that this is ascribed to the presence of the
large-scale filamentary structures in those models which the large
$\theta_{12}$ (therefore small $|{\mb k}_1+{\mb k}_2|$) in $k$-space may
correspond to.

The bottom panels in Figure \ref{fig:phasesum_scalefree} show the
dependence on the sampling volume size $V_{\rm samp}$. Again unlike
the conventional statistics, the PDF of the phase sum is indeed
sensitive to the choice of $V_{\rm samp}$. The lower-right panel of
Figure \ref{fig:p3_scalefree} illustrates that $p^{(3)} \sqrt{V_{\rm
samp}}$ is approximately constant implying the validity of the
hierarchical clustering ansatz as discussed in \S 2. This is why
non-Gaussian feature of the PDF becomes substantially stronger and the
higher-order terms become more important as $V_{\rm samp}$ becomes
smaller.

\begin{figure*}[tph]
\begin{center}
\epsfxsize=16cm 
\plotone{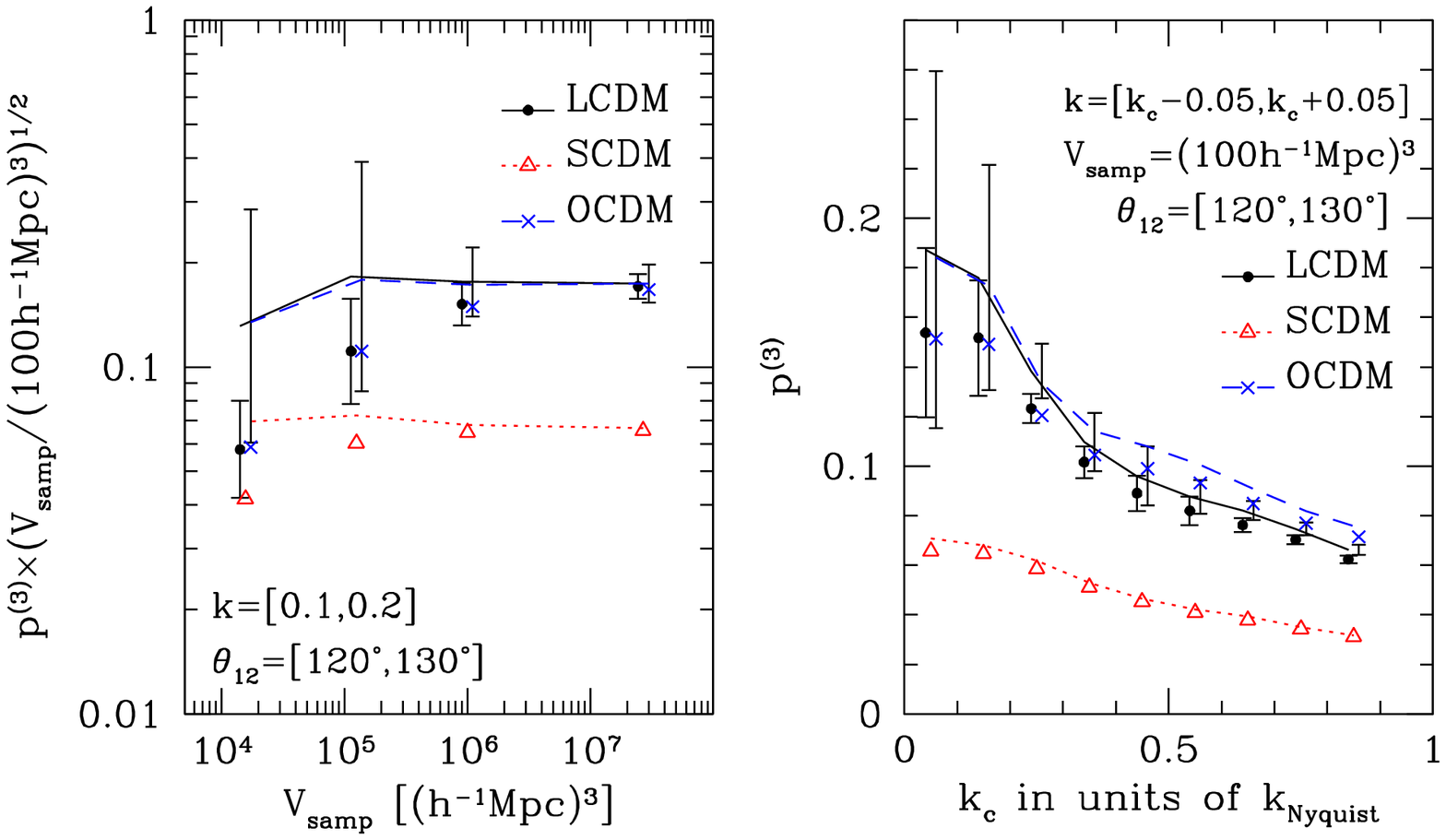}
\end{center}
\caption{Box-size and scale dependence of $p^{(3)}$ (symbols) in comparison
with the amplitude of the distribution of the phase sum 
at $\theta_{{\smb k}_1}
+\theta_{{\smb k}_2}-\theta_{{\smb k}_1+{\smb k}_2}=0$ normalized by
the constant factor $\pi^{3/2}/4$ (lines) for CDM models.
The error bars indicate the sample variance of $p^{(3)}$ and
the amplitude of the distribution of the phase sum for LCDM model.
Each panel shows the dependence of $V_{\rm samp}$ ({\it Left}) 
and $|{\mb k}_1|(|{\mb k}_2|)$ in units of 
$k_{\rm Nyquist}=2\pi/(0.78h^{-1}{\rm Mpc})$ ({\it Right}).
\label{fig:p3_cdm}}
\end{figure*}

Figures \ref{fig:phasesum_cdm} and \ref{fig:p3_cdm} plot the same
results as Figures \ref{fig:phasesum_scalefree} and
\ref{fig:p3_scalefree} but for the CDM models. The distribution of the
phase sum is more uniform in SCDM models compared with other models
mainly due to the smaller value of $\sigma_8$ (Table
\ref{tab:modelpara}). This is because the small $\sigma_8$ corresponds
to the small power $P(k)$ with all over the scales and then
$p^{(3)}\sim\sqrt{P(k)/V}$ has low value. The behavior of the
parameter dependence is almost same as that in scale-free models. When
a length of the simulation box is $300h^{-1}$Mpc, the lowest order
approximation well agrees with simulations for any CDM models, while
the approximation becomes worse as the box length decreases.

\begin{figure*}[tph]
\begin{center}
\end{center}
\caption{Comparison between the distribution of phase sum 
$\theta_{{\smb k}_1}+\theta_{{\smb k}_2}-\theta_{{\smb k}_1+{\smb k}_2}$ 
plotted by symbols and the lowest order perturbative equation
(eq.[\ref{eq:matsubara}]) plotted
by lines for CDM models. The sampling volume $V_{\rm samp}$
is $(300h^{-1}{\rm Mpc})^3$, $(100h^{-1}{\rm Mpc})^3$, 
$(50h^{-1}{\rm Mpc})^3$, and $(25h^{-1}{\rm Mpc})^3$ from top to bottom. 
The range of the modulus $|{\mb k}_1|$ and $|{\mb k}_2|$ is 
$[0.1,0.2]$(open circles and solid lines), 
$[0.4,0.5]$(closed triangles and dotted lines), 
and $[0.8,0.9]$(crosses and dashed lines) in units of
the Nyquist wavenumber $k_{\rm Nyquist}$ written in each panel. 
\label{fig:phasesum_cdm}}
\end{figure*}

\subsection{Correspondence between phase correlations and density peaks}

As we emphasized over and over again, the PDF of the phase sum does
depend on the size of the sampling volume. In order to understand the
physical correspondence between the real-space distribution and the
phase correlation, we compare the particle distribution from small
sampling volumes on a sample-to-sample basis.

First note that the non-uniformity of the phase sum arises from a few
dominating density peaks in the sampling volume.  Consider a field
which has just one spherical peak with a density profile $h(r)$
located at $\mb{x}_0$. The density field in this idealized system is
given by $\rho(\mb{x}) = h(|\mb{x}-\mb{x}_0|)$, and irrespective of
the density profile of the peak, all the Fourier phases are aligned to
$\theta_{\smb{k}} = -\mb{k}\cdot\mb{x}_0$. Thus the phase sum becomes
zero in any combination of closed wavevectors.  If another peak exists
at a different position, the alignment of the phases is broken
depending on the relative strength of the two peaks.  If the peak
height of the second peak is much smaller than that of the first, the
phase alignment is almost preserved and the phase sum remains
distributed around zero.  On the other hand, if they have a comparable
peak height, the phase alignment is significantly broken.  In reality,
all density peaks should contain substructure to some extent and many
other structures in the field also contribute the eventual phase
distribution. Nevertheless the above qualitative picture proves to be
useful in understanding the physical meaning of the phase sum
distribution.

Motivated by the above consideration, we examine in detail the
connection between the particle distribution and the PDF of the phase
sum.  Figure \ref{fig:field_vs_phase_25Mpc} plots the projected
particle distribution and the corresponding PDF of the phase sum for 8
different samples of ($25h^{-1}$Mpc)$^3$ box from one of the LCDM
simulations of ($100h^{-1}$Mpc)$^3$ box.

After smoothing the particle distribution over the Gaussian window of
$1h^{-1}$Mpc radius, we compute genus using CONTOUR 3D
\cite{Weinberg1988}.  Then we define the peak heights of the 1st, 2nd
and 3rd peaks, $\delta_1$, $\delta_2$, and $\delta_3$ by the threshold
density contrasts which the number of genus in the box changes from
from $-1$ to $0$, from $-2$ to $-1$ and from $-3$ to $-2$,
respectively. Those values are indicated in each panel of Figure
\ref{fig:field_vs_phase_25Mpc}. As expected from the above argument,
Figure \ref{fig:field_vs_phase_25Mpc} clearly shows that the degree of
non-uniformity of the PDF is related to the presence of the dominating
peak in the box; the sample (d) has a single prominent density peak,
and the phase sum is strongly distributed around zero, while the
sample (b) has three almost comparable peaks resulting a fairly
uniform PDF of the phase sum.

\begin{figure*}[tph]
\begin{center}
\epsfxsize=16cm 
\plotone{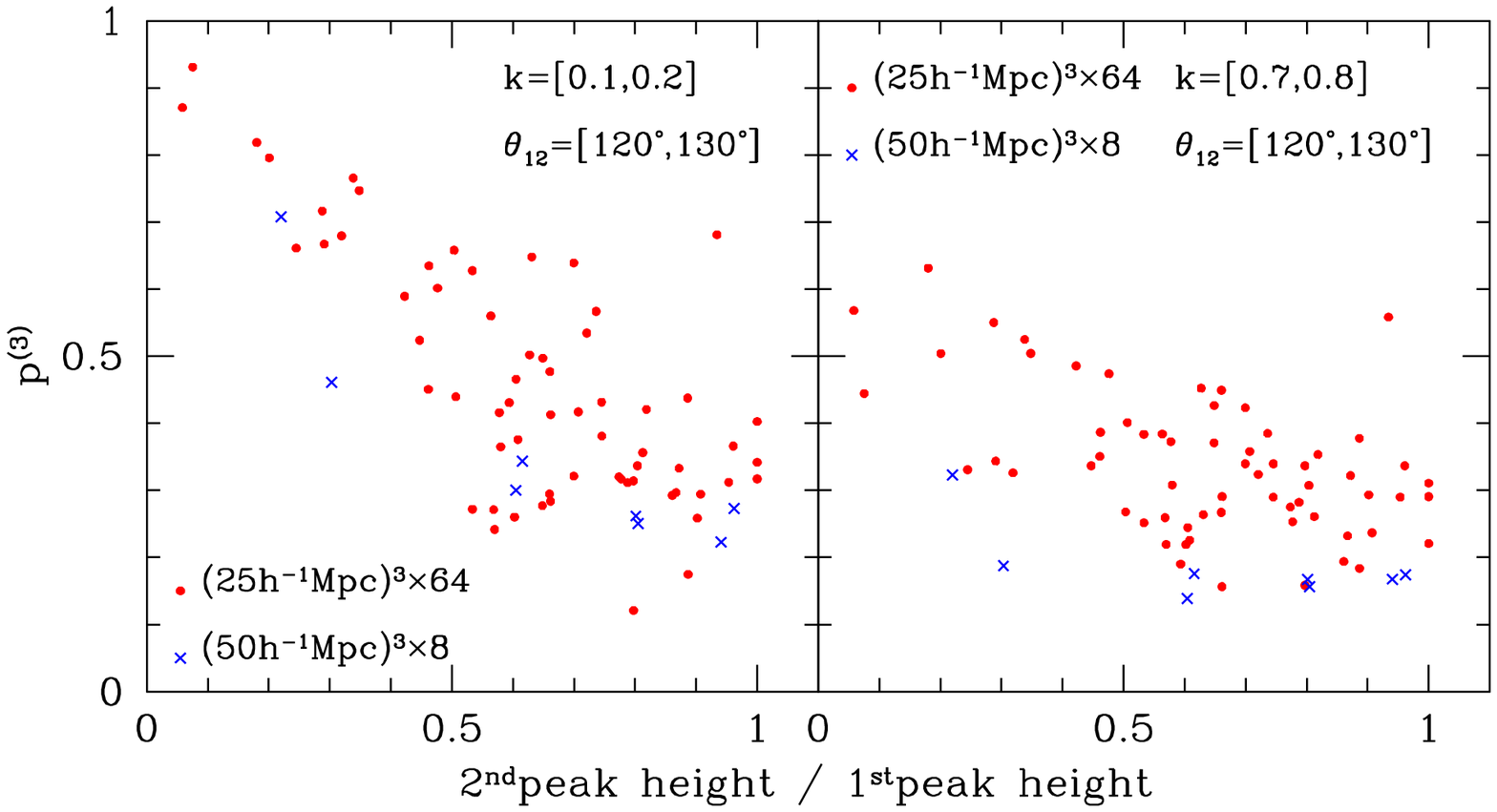}
\end{center}
\caption{The relation between $p^{(3)}$ and the ratio of the second
highest peak density to the highest peak density for the $64$ cubic
data with a box-length of $25h^{-1}$Mpc (filled circles) and the $8$
cubic data with a box-length of $50h^{-1}$Mpc (crosses) based on LCDM
cosmological model.  The ranges of scale $|{\mb k}_1|$ and $|{\mb
k}_2|$ are $[0.1,0.2]$ ({\it Left}) and $[0.7,0.8]$ ({\it Right}) in
units of $k_{\rm Nyquist}=2\pi/(0.78h^{-1}{\rm Mpc})$ and
$\theta_{12}=[120^\circ,130^\circ]$.
\label{fig:peakratio_vs_p3}}
\end{figure*}

To be more quantitative, we compute $p^{(3)}$ and the ratio of the
peak heights of the first and the second peaks for $64$ samples of
$V_{\rm samp}=(25h^{-1}$Mpc)$^3$ and $8$ samples of $V_{\rm
samp}=(50h^{-1}$Mpc)$^3$, both constructed from one realization of
LCDM model. The result plotted in Figure \ref{fig:peakratio_vs_p3}
indeed supports the strong correlation between the value of $p^{(3)}$
and the ratio of the peak heights. In fact, this naturally explains
the dependence of the PDF on $V_{\rm samp}$. As $V_{\rm samp}$
increases, the number of density peaks in the sampling volume
increases and those peaks tend to weaken the correlation of the phase
sum. This clearly illustrated in Figure \ref{fig:field_vs_phase_50Mpc}
which plots the result for the box of $V_{\rm samp}=(50h^{-1}$Mpc)$^3$
comprising all the eight samples of $V_{\rm samp}=(25h^{-1}$Mpc)$^3$
plotted in Figure \ref{fig:field_vs_phase_25Mpc}.

\begin{figure*}[tph]
\begin{center}
\end{center}
\caption{Comparison of the distribution of the phase sum with the
value of three highest density peaks for eight fields with a
box-length of $25h^{-1}$Mpc. The projected maps of dark matter
particles are shown together.  The distributions of the phase sum
(symbols) are plotted together with the corresponding lowest order
perturbative formula (lines) at three ranges of scale $|{\mb k}_1|$
and $|{\mb k}_2|$ in $[0.1,0.2]$ (filled circles, solid lines),
$[0.4,0.5]$(open triangles, dotted lines) and $[0.7,0.8]$ (crosses,
dashed lines) in units of $k_{\rm Nyquist}=2\pi/(0.78h^{-1}{\rm
Mpc})$.
\label{fig:field_vs_phase_25Mpc}}
\end{figure*}

\begin{figure*}[tph]
\begin{center}
\end{center}
\caption{Same as Figure \ref{fig:field_vs_phase_25Mpc} but for a box
with a length of $50h^{-1}$Mpc constituted from the eight sub-boxes
shown in Figure \ref{fig:field_vs_phase_25Mpc}.
\label{fig:field_vs_phase_50Mpc}}
\end{figure*}

\section{Conclusions and Discussion}

We have performed a detailed numerical analysis of the phase
correlation induced by nonlinear gravity in the universe. Following
recent analytic work by Matsubara (2003b) in perturbation theory, we
have computed the distribution function of the phase sum in triangle
wavevectors and compared with his analytic formula. Using a large set
of $N$-body simulations, we have explored the behavior of phase
correlations for various configurations of wavevector triangles in the
Fourier space at different epochs in a range of different sampling
volumes. We find that the agreement of the simulations and the
analytic perturbation formula is generally good.

Quite interestingly the series expansion of the analytic formula
breaks down for $P(k)/V_{\rm samp}\sim 1$, which is in marked contrast
to the conventional linear/nonlinear criterion, $P(k)k^3 \sim 1$.
Therefore the statistics of the phase sum indeed depends on the size
of the sampling volume explicitly.  This feature is naturally
explained by the corresponding density structure in real space; all
the phases are synchronized with each other when a distinctive density
peak dominates in the given sampling volume, while the phase
correlations become weaker if several peaks with comparable heights
exist.  We have confirmed this expectation directly by comparing the
strength of phase correlations with the ratio of the highest peak to
the secondary peaks in the sampling volume. Therefore the phase
correlation becomes diluted as the sampling volume increases because
statistically it accommodates a number of peaks with comparable
heights.

The fact that the perturbation formula well reproduces the numerical
results indicates that the information from the phase sum distribution
mainly comes from the normalized bispectrum amplitude $p^{(3)}$.
Since one can evaluate the bispectrum directly and independently from
the phase sum distribution, one might wonder if the phase sum
statistics is simply equivalent to the more conventional bispectrum
analysis. Indeed the dependence on the sampling volume size may be a
key in this context.  This implies that any signal of non-Gaussianity
in the phase correlation disappears in the large-volume limit. We can
repeat the analysis by systematically decreasing the sampling volume
size.  Instead, one has to take an ensemble average of the phase
correlations over many subsamples of a fixed $V_{\rm samp}$ to obtain
proper statistical quantities. Otherwise, the phase correlations in a
single volume is much like the cosmic variance, as the sharpness of
the highest peak is weakened in the limit $V_{\rm samp} \rightarrow
\infty$. The situation is similar to the count-in-cells statistics, as
the exact number of the galaxies in a single cell is attributed to the
cosmic variance. Taking an average of the weighted number in many
cells gives a proper statistics, such as variance, skewness, kurtosis,
and so forth.  The average value and the sample variance with respect
to the average as a function of $V_{\rm samp}$ may give additional
information even beyond the bispectrum.

The distribution of the phase sum may be applied to the real data of
the existing galaxy catalogues including the 2dF and SDSS, which
provides complementary information beyond the two-point statistics.
Since the phase sum statistics is a new concept in cosmology, we still
need to explore many other aspects of the statistic that we are likely
to have overlooked at this point.  We are currently working on
evaluating the phase-sum distributions of Sloan Digital Sky Survey
galaxy samples considering all possible systematics by creating
simulated Mock samples as we did in the previous topological analysis
(Hikage et al. 2002, 2003).  We hope to report the results elsewhere
in near future.

\acknowledgments

We deeply appreciate I. Kayo and Y. P. Jing for kindly providing a
large set of $N$-body simulation data.  We thank K. Yoshikawa for
helping us improve our numerical analysis routines.  C.H. also thanks
G. Watanabe, and A. Taruya for useful discussion.  Numerical
computations were carried out at ADAC (the Astronomical Data Analysis
Center) of the National Astronomical Observatory, Japan (project ID:
yys08a). This research was supported in part by the Grant-in-Aid from
Monbu-Kagakusho and Japan Society of Promotion of Science (12640231,
14102004, 1470157 and 15740151).


\clearpage
\appendix

\section{Box-size dependence of the cumulants $p^{(N)}$}
\label{app:extensivity}

According to the analytic formula of the joint PDF of arbitrary sets
of the Fourier phase $\theta_{\smb k}$, the distribution of the phase
sum is determined by the cumulants $p^{(N)}=\langle \alpha_1\alpha_2
\cdots \alpha_N \rangle_c$ where $\alpha_i =f_{{\smb
k}_i}/\sqrt{\langle|f_{{\smb k}_i}|^2\rangle}$ ($N\ge 3$ and
$\langle\cdots\rangle$ denotes the spatial average). In this section,
we present the relation of the cumulants $p^{(N)}$ to the power
spectra and the polyspectra in physical unit (independent on the
box-size) and explicitly show the dependence of $p^{(N)}$ on the
box-size. First we give the definition of the power spectrum and the
polyspectra and then apply it to the grid data.

In a three dimensional density fluctuation field $\delta({\mb
x})=(\rho({\mb x})-\bar{\rho})/\bar{\rho}$, the Fourier-transform
$\tilde\delta({\mb k})$ of $\delta({\mb x})$ in an infinite volume is
given by
\begin{equation}
\tilde{f}({\mb k})=\int d^3x\delta({\mb x})e^{-i{\smb k}\cdot{\smb x}}.
\end{equation}
The polyspectra $P^{(N)}({\mb k}_1,\cdot\cdot\cdot,{\mb k}_{N-1})$ 
are given by the cumulants
\begin{equation}
  \left\langle \tilde{f}({\mb k}_1) \cdots  \tilde{f}({\mb k}_N)
  \right\rangle_c = (2\pi)^3
\delta_D^3({\mb k}_1+\cdot\cdot\cdot{\mb k}_N) P^{(N)}({\mb k}_1,
\cdot\cdot\cdot,{\mb k}_{N-1}),
\end{equation}
where $\delta_D(\mb k)$ is Dirac's delta function. In this definition,
the variance $\sigma^2 = \langle\delta^2({\mb x})\rangle$ is given by
\begin{equation}
  \sigma^2= \int \frac{k^2 dk}{2\pi^2} P(k),
\end{equation}
where $P(|{\mb k}|)=P^{(2)}({\mb k})$ is the conventional power
spectrum.

In a simulated box with a volume $V$ and a
total grid number $N$, the lattice spacing of
${\mb x}$ is $(V/N)^{1/3}$. In a lattice field
$\delta({\mb x})$ with a discrete set of $x$'s, the discrete Fourier
transform $f_{\mb k}$ is computed by
\begin{equation}
  f_{\smb k}=\frac{\sqrt{V}}{N}
  \sum_{\smb x}\delta({\mb x}) e^{-i{\smb k}\cdot{\smb x}},
\end{equation}
where the lattice spacing in $k$-space is $2\pi/V^{1/3}$. The
correspondence between the discrete Fourier modes and the continuum
Fourier modes is
\begin{equation}
   \lim_{V\rightarrow\infty, V/N \rightarrow 0}
  \sqrt{V} f_{\smb k} = \tilde{f}({\mb k}).
\end{equation}
The cumulants $\left\langle f_{{\smb k}_1} \cdot\cdot\cdot f_{{\smb
k}_N} \right\rangle_c$ are related to $P^{(N)}({\mb k}_1,
\cdot\cdot\cdot , {\mb k}_{N-1})$ by the relation
\begin{equation}
\left\langle f_{{\smb k}_1} \cdot\cdot\cdot f_{{\smb k}_N} \right\rangle_c
 =  V^{1-N/2} \delta^K_{{\smb k}_1+ \cdots +{\smb k}_N}
P^{(N)}({\mb k}_1, \cdot\cdot\cdot ,{\mb k}_{N-1}),
\end{equation}
where $\delta^K_{\smb k}$ is Kronecker's delta which is $1$ $({\mb
k}=0)$ and $0$ $({\mb k}\ne 0)$. The correspondence between the
Kronecker's delta and the Dirac's delta function is
\begin{equation}
   \lim_{V\rightarrow\infty, V/N \rightarrow 0}
   V \delta^K_{\smb k} = (2\pi)^3 \delta_D^3({\mb k}).
\end{equation}

The cumulants $p^{(N)}$ are, therefore, calculated by
\begin{eqnarray}
p^{(N)} & = & \frac{\left\langle f_{{\smb k}_1}\cdot\cdot\cdot f_{{\smb k}_N} 
\right\rangle_c}{\sqrt{\langle|f_{{\smb k}_1}|^2\rangle
\cdot\cdot\cdot\langle|f_{{\smb k}_N}|^2\rangle}} \nonumber \\
&=& \frac{P^{(N)}({\mb k}_1, \cdot\cdot\cdot ,{\mb k}_{N-1})}{\sqrt{V^{N-2}
P({\mb k}_1)\cdot\cdot\cdot P({\mb k}_N)}}.
\end{eqnarray}
The box-size dependence of $p^{(N)} \propto V^{1-N/2}$ means that
higher order terms become dominant as the box-size decreases even
when $k^3P(k)$ is small.


\begin{thebibliography}{}
\parskip=-1pt
\baselineskip=14pt

\bibitem[Bardeen et al. 1986]{BBKS1986}
Bardeen, J. M., Bond, J. R., Kaiser, N., \& Szalay, A. S.
1986, ApJ, 304, 15

\bibitem[Bertschinger 1992]{Bertschinger1992}
Bertschinger, E., in Lecture Notes in Physics, 408,
{\it New Insights into the Universe}, ed. Martinez, V. J., 
Portilla, M., \& Saez, D., p65 (Springer-Verlag, Berlin, 1992)

\bibitem[Chiang 2001]{C2001}
Chiang, L. 2001, MNRAS, 325, 405

\bibitem[Chiang et al. 2002]{CCN2002}
Chiang, L., Coles, P., \& Naselsky, P. D. 2002, MNRAS, 337, 488

\bibitem[Coles \& Chiang 2000]{CC2000}
Coles, P., \& Chiang, L. 2000, Nature, 406, 376

\bibitem[Gott et al. 1986]{GMD1986}
Gott III, J. R., Mellot, A. L., \& Dickinson, M. 1986, ApJ, 306, 341

\bibitem[Hikage et al. (2002)]{H2002} Hikage,~C., Suto,~Y., Kayo,~I.,
Taruya,~A., Matsubara,~T., Vogeley,~M.~S., Hoyle,~F., Gott,~J.~R.,~III,
\& Brinkmann,~J., for the SDSS collaboration 2002, \pasj, 54, 707

\bibitem[Hikage et al. (2003)]{H2003} Hikage,~C., Schmalzing, J.,
Buchert, T., Suto,~Y., Kayo,~I., Taruya,~A., Vogeley,~M.~S., Hoyle,~F., 
Gott,~J.~R.,~III, \& Brinkmann,~J., for the SDSS collaboration 2003, 
\pasj, 55, in press.

\bibitem[Jain \& Bertschinger 1998]{Jain1998}
Jain, B., \& Bertschinger, E. 1998, ApJ, 509, 517

\bibitem[Jing 1998]{Jing1998}
Jing, Y. P. 1998, ApJ, 503, L9

\bibitem[Jing \& Suto 1998]{JS1998}
Jing, Y. P., \& Suto, Y. 1998, ApJ, 494, L5

\bibitem[Matsubara 1995]{Matsubara1995}
Matsubara, T. 1995, ApJS, 101, 1

\bibitem[Matsubara 2003a]{Matsubara2003a}
Matsubara, T. 2003a, ApJ, 584, 1

\bibitem[Matsubara 2003b]{Matsubara2003b}
Matsubara, T. 2003b, ApJL, 591, L79

\bibitem[Mecke et al. 1994]{MBW1994}
Mecke, K. R., Buchert, T., \& Wagner, H. 1994, A\&A, 288, 697

\bibitem[Ryden \& Gramann 1991]{Ryden1991}
Ryden, B. S., \& Gramann, M. 1991, ApJ, 383, L33

\bibitem[Scherrer et al. 1991]{SMS1991}
Scherrer, R. J., Melott, A. L., \& Shandarin, S. F. 1991, ApJ, 377, 29

\bibitem[Schmalzing \& Buchert 1997]{SB1997}
Schmalzing,~J., \& Buchert,~T.\ 1997, ApJ, 482, L1

\bibitem[Soda \& Suto 1992]{Soda1992}
Soda, J., \& Suto, Y. 1992, ApJ, 396, 379

\bibitem[Suginohara \& Suto 1991]{Suginohara1991}
Suginohara, T., \& Suto, Y. 1991, ApJ, 371, 470

\bibitem[Watts \& Coles 2003]{WC2003}
Watts, P., \& Coles, P. 2003, MNRAS, 338, 806

\bibitem[Watts et al. 2003]{WCM2003}
Watts, P., Coles, P., \& Melott, A. 2003, ApJ, 589, L61

\bibitem[Weinberg 1988]{Weinberg1988}
Weinberg, D. H. 1988, PASP, 100, 1373

\bibitem[White 1979]{White1979}
White, S. D. M. 1979, MNRAS, 186, 145
\end{thebibliography}
\end{document}